# Can we advance macroscopic quantum systems outside the framework of complex decoherence theory?

Mark E. Brezinski[1,2,3] and Maria Rupnick [1,2]

[1] Center for Optical Coherence Tomography and Modern Physics, Department of Orthopedic Surgery, Brigham and Women's Hospital, 75 Francis Street, MRB-114, Boston, MA 02115.

[2] Harvard Medical School, 25 Shattuck Street, Boston, MA 02115.

[3] Department of Electrical Engineering and Computer Science, Massachusetts Institute of Technology, Rm 36-360, 50 Vassar St., Cambridge, MA 02139.

Corresponding author:
Mark E. Brezinski, MD, PhD
Department of Optical Coherence Tomography and Modern Physics
Harvard Medical School
Massachusetts Institute of Technology
MRB-114
75 Francis Street
Boston, Mass. 02115
Tel: (617)233-2802
Email address:  mebrezin@mit.edu

Macroscopic quantum systems (MQS) are macroscopic systems driven by quantum rather than classical mechanics, a long studied area with minimal success till recently. Harnessing the benefits of quantum mechanics on a macroscopic level would revolutionize fields ranging from telecommunication to biology, the latter focused on here for reasons discussed. Contrary to misconceptions, there are no known physical laws that prevent the development of MQS. Instead, they are generally believed universally lost in complex systems from environmental entanglements (decoherence). But we argue success is achievable MQS with decoherence compensation developed, naturally or artificially, from top-down rather current reductionist approaches. <u>This paper advances the MQS field by a complex systems approach to decoherence.</u> First, why complex system decoherence approaches (top-down) are needed is discussed. Specifically, complex adaptive systems (CAS) are not amenable to reductionist models (and their master equations) because of emergent behavior, approximation failures, not accounting for quantum compensator mechanisms, ignoring path integrals, and the subentity problem. In addition, since MQS must exist within the context of the classical world, rapid decoherence and prolonged coherence are both needed. Nature has already demonstrated this for quantum subsystems such as photosynthesis and magnetoreception. Second, we perform a preliminary study that illustrates a top-down approach to *potential* MQS. In summary, reductionist arguments against MQS are not justifiable. It is more likely they are not easily detectable in large intact classical systems or has been destroyed by reductionist experimental set-ups. This complex systems decoherence approach, using top down investigations, is critical to paradigm shifts in MQS research both in biological and non-biological systems.

# Part I: General Introduction

Consistently harnessing the benefits of quantum mechanics on a scale above the microscopic level would revolutionize fields ranging from telecommunication to biology. Macroscopic/mesoscopic quantum systems (MQS) are then the driving of large scale or complex systems by quantum rather than classical mechanics. It is area of investigation stretching back to the origins of quantum mechanics, though with minimal success till recently [1-23].

There are no known physical laws that prevent the development of MQS that we as well as others have recently addressed [9-19]. The common view for their lack of observation is coherence loss from overwhelming environmental entanglements (decoherence) [24-28]. Both these concepts are reviewed in part II. But no understanding of decoherence in terms of complex systems currently exists, making it challenging to identify or create MQS, which can be unobserved in the classical environment. Current decoherence theory and the associated master equations (used for modeling and experimental design) are reductionist, attempting to extrapolate data from atomic and subatomic systems to complex systems. This paper's focus is to demonstrate this widely used reductionist approach to complex systems decoherence are generally doomed to failure. Instead we strongly support a complex systems (top-down) approach modeled by finding equivalent classes and transition functions that result in the development of homomorphisms. '

In part II, because this paper involves the overlap of numerous distinct disciplines, several topics will be reviewed. This again includes the quantum principles of coherence and decoherence in small particle systems. They both will be examined in terms of a simple Young's Interferometer and then density operator formalism. The density operator formalism in particular is important because it will allow master equations to be introduced (with their limitations discussed in part III).

In part II, we will also discuss the concepts of a complex and more importantly complex adaptive systems (CAS). Since the most numerous examples of CAS are biological systems, much of the focus will be on them, though by no means are the results of this paper intended to be limited to quantum biology. The 'trivial' quantum mechanics of biological molecular bonds is well known, but are not the subject of this paper [68-71]. Non-trivial examples that are discussed in the paper extensively are the quantum subsystems of magnetoreception and photosynthesis. Macroscopic quantum advances have been achieved in non-biological systems. Examples include SQUID devices, remote mirror correlations, Bose-Einstein condensate, optics, and insights into the quantum computer (albeit under extreme conditions)[1-19]. Optics examples include work by our group where recently we have demonstrated non-local macroscopic quantum correlations established between remote reflectors, under ambient conditions [135]. Similarly, other groups have achieved quantum correlations with diamonds or polarization states, with all three advancing the field of MQS [136-137]. But biological systems have the advantage of evolutionary adaptation for developing decoherence compensation. This is an extension of the dictum that whatever technology humans invent; nature normally gets there first and more effectively. This can be seen, for example, with the high-energy efficiencies in photosynthesis and sensitivities of magnetoreception greater than any man made device (discussed in appendix).

Also from these two subsystems, for quantum mechanics to play a non-trivial role, coherence must last durations much longer than suggested by native thermal (wet) reductionist calculations (master equations). This will be discussed in part II. On the other hand, rapid decoherence is vital for bestowing upon molecules well-defined three-dimensional shapes (as opposed to superpositions). For the majority of cells or device functions, classical geometry and topology are still crucial, separate from any quantum phenomena. So, it appears that both rapid decoherence and prolonged coherence need to co-exist within complex systems. In other words, ambient systems need to exploit the best of both worlds—decoherence and coherence, far more complicated than reductionist models. This also makes it challenging to identify quantum systems in classical environments.

Then, beginning in part III, the theoretical basis of why a complex systems approach to decoherence is essential is discussed in extensive detail. We will argue that using reductionist approaches (and their modeling by master equations) to extrapolate to large-scale quantum systems is futile. So making statements that quantum phenomena are prohibited from being sustained at a macroscopic scale, because of extrapolated reductionist decoherence data, is not justifiable. It is just as likely that they are not easily detectable in the context of the large intact classical systems or destroyed by partially isolating the system during experiments.

The core of the paper, the failure of reductionist approaches for modeling decoherence in CAS, are discussed in part III. Reductionism ignores the higher order behaviors of CAS, such as emergence, compartmentalization, robustness, and broad adaptability that are lacking in simpler systems [47-59].   In addition, the reductionist approach does not take into account other properties of complex systems and do not allow calculations from master equations to be extrapolated to complex systems.  The first is the essentially universal use of approximations (Born and Markov most commonly) in decoherence master equations that would not be valid for CAS.  The second is not taking path integrals into account in master equations.  The third is not addressing decoherence avoidance or compensation mechanisms, even those already demonstrated with man-made systems (such as in the quantum information sciences). The fourth is not addressing the subentity problem, particularly in the use of the trace of density and time evolution operators in master equations.  Therefore, reductionist master equations are not representative of complex systems, so the difficulty in demonstrating MQS using theory based on them is not surprising.

Finally, in part IV, we reduce the theory to practice.  Preliminary data is generated that illustrates the top-down approach to evaluate a potential quantum system.  We had previously seen anecdotal evidence of non-local correlations in HL-60 cells, as discussed below.  We postulate that this is secondary to quantum correlations, and want to exclude biomolecule exchange or direct contact.  While more studies need to be done, it is at least consistent with a non-local remote quantum system.

It should be noted that there is ambiguity in the literature as to what is a macroscopic versus mesoscopic quantum system.  Are the magentoreceptor or photosynthesis subsystems, which drive entire birds or plants, mesoscopic or macroscopic[62]?  Some authors say yes while others say no.  Similarly, if in theory two molecules are non-locally (remote in classical space) quantum correlated in spin, is it mesoscopic or macroscopic?  With ambiguity in the literature, in this paper we will use the terms mesoscopic and macroscopic quantum systems (MQS) as synonymous.

## Part II:  The Reductionist Approach to Coherence and Decoherence

Before addressing the limitations of reductionist decoherence theory and their master equations to complex systems (part III), this section will review the reductionist approach (and master equations) to decoherence both in terms of a Young's interferometer and then with density operator formalism.  Also, we will give the general description why, once maintained, a coherent state remains coherent unless there is an interaction.  Finally, in this sections will we review the CAS, needed in part III.

### IIA. Coherence and Decoherence Illustrated With a Simple Young's Experiment

A convenient illustrative way to express a coherent state is in terms of path indistinguishability with a Young's interferometer.  We will begin using a simple model (reductionist) with and without decoherence from pioneers like Zurek and Zeh, where local environmental entanglements lead to path distinguishability (loss of coherence) as well as non-locality [25-28].  This is insightful work for a simple system, but unfortunately it is the widely held view this is also how decoherence operates on the macroscopic level.  We will demonstrate why this is not the case in part III.  Then from this relatively simple model we will be re-expressing decoherence in terms of the density operator.

As a simple example of the relationship between decoherence and indistinguishable paths (as well as non-locality), consider the Young's interferometer in figure 1.  Here the source is a beam, with particles entering the interferometer one at a time that makes it easier to illustrate non-locality.  This interferometer has a barrier limited to two slits that reduces the number of high probability paths.  For purposes of the current discussion we will assume the barrier, with two slits, reduces it to two paths (i.e ignoring the path/field integral formulation of many paths through the two slits for now).  Later we will extend this further by addressing how environmental interactions (influential functions in the path integral formulation) can potentially be sculpturing the action (path integral) but not completely eliminating coherence/interference.  In the Young's set-up, in the absence of decoherence, the paths are completely indistinguishable and for coherent particles, interference results on the screen behind (first order coherence).  The interference pattern, at the screen, is given by the expanded density operator:

$$\hat{\rho} = \frac{1}{2}\left\{|\psi_1\rangle\langle\psi_1| + |\psi_2\rangle\langle\psi_2| + |\psi_1\rangle\langle\psi_2|\langle E_2|E_1\rangle + |\psi_2\rangle\langle\psi_1|\langle E_1|E_2\rangle\right\} \quad (1)$$

In this description, E in the equation will represent the environmental interactions/entanglements. Reductionist theory treats these interactions as irreversible but we will see may be reversible or in some special cases expand coherence.  The event that occurs at the screen is measurement, where a single eigenvalue is selected. In this equation, 1 and 2 correspond to the two potential paths (and only two) the particle can take and ψ is the wavefunction for the particle. If we initially ignore the E terms (environmental interactions/entanglements), the pattern of interaction on the screen (measurement) demonstrates the interference pattern that is represented by the last two terms in the equation (while the first two terms represent DC terms). Now, if $E_1$ and $E_2$ are substantially different terms (inner product near zero), the third and fourth terms disappear. Interference (coherence) is lost in this simplistic example of environmentally induced decoherence (which occurs through the development of path distinguishability) that follows the Zurek and Zeh approach. The amount of difference between the E terms (environmental entanglement terms) affects the degree to which coherence (and interference) is lost. If $E_1$ and $E_2$ are similar (inner product 1), the paths are indistinguishable even though entanglements/correlations with the environment occurred, and the interference pattern is maintained. Therefore, unlike reductionist approaches, the environmental entanglements do not necessarily lead to decoherence when the E terms are nearly identical. In two previous papers we expand on this and describe how coherence of the system could actually be expanded (a closed system to a larger closed system) with environmental entanglements under the proper conditions [22,23]. Other groups, but using reductionist systems far from ambient conditions, have also described this [10,14,19].

   An insightful experimental modification, that will become relevant to decoherence compensation later, is the results which occur if the screen is placed at A, B, or C in the figure (the environmental entanglements here are identical). If the screen is placed in either the A or C positions an interference pattern will result, but when in position B interference is lost because the paths are distinguishable. And if the screen is moved from B to C, the interference pattern is recovered and decoherence reversed. This is similar to quantum erasers and time reversals [75,76]. It is one strategy we will examine for dealing with decoherence in complex systems: its reversibility (unlike measurement). The key aspect (in maintaining coherence/correlations) is whether the two paths are distinguishable/indistinguishable <u>when a measurement is ultimately performed.</u> Prior to measurement, potentials are added, while after measurement, intensities are added. It is clear that measurement is occurring at the screen, with the particle irreversibly interacting with the screen establishing one eigenvalue. However, we note several important points of environmental entanglements relative to decoherence theory in this <u>simple</u> model. The first is that if the interactions are similar in both paths, the coherence is maintained, and in general lost if they are different. The second is that under some conditions decoherence is reversible, as illustrated. Third, the interaction with the screen represents measurement, an irreversible process distinct from decoherence and leading for a single eigenvalue for one particle. The fifth is that single particle interference is non-local. What happens on the screen is effected by what available paths exist for the single photon. The fifth is the two paths in this simple example are obviously a subset of the broader concept of path integrals (the action), which can have an infinite number of paths. This too will be discussed under failure of reductionist models. We will discuss below how these paths may interact with the environment differently, which may or may not alter the action. Master decoherence equations rarely take into account the path/field integrals and essentially never consider the fact that the action may be maintained in a wet environment (when some but not all paths are lost), particularly in a CAS. Six, under the proper conditions, environmental entanglements can lead to expansion rather than loss of coherence (going from a closed system to another larger closed system). This occurs under specific conditions which we have discussed elsewhere, but an inner product near one, as described in equation 1, is an important component from our previous work [22,23], as well as violating both the Bornian and Markovian approximations.

IIB. Closed and Opened Quantum Systems

   The previous section illustrated coherence and the reductionist model of decoherence using single particles in an interferometer. Coherence is lost as a function of particle types of interactions. Here we expand that to include an indefinite number of particles and do so using the density operator formalism. It will illustrate why quantum mechanics does not prohibit the existence of MQS and why decoherence compensation remains the critical component to success in the MQS field. The properties of a density operator will be briefly reviewed as it is used in the remainder of the paper. The density operator is a Hermitian operator acting on Hilbert space with nonnegative eigenvalues whose sum is 1. It is an 'operator <u>on</u> Hilbert space' but often is treated as a classical statistical operator, which it is not. This distinction will become obvious when we discuss the subentity problem.

The density operator (state matrix or density matrix) has the advantage over the wavefunction (in representing a state) in that it can describe mixtures. A pure state is a one dimensional projection operator, follows the condition $p^2 = p$, and completely defines a system (no ignorance interpretation). When a single vector in Hilbert space can describe a quantum system, we are dealing with a pure state, and if not, a mixed state exists (the interpretation of a mixed state without measurement is discussed below, as some consider it is also a pure state until measurement).

A density operator does not specify a unique microscopic classical configuration (nor can anything else), which is not surprising based on its definition, but it is sufficient for calculating the expectation values of physical properties. The density operator also contains the information about superpositions between subsystems. The expectation value is generally given by the trace of the product of the density operator and the observable [77,78]:

$$\{\langle |O| \rangle\}_{expec} = \text{Tr}\{\rho O\} \qquad (2)$$

This point will be raised again later, but by obtaining the expectation value through the trace, we are eliminating off-diagonal elements and therefore coherences. It represents the expectation value of a near infinite number of measurements at that point (which is a real number). It does not typically yield a specific state, but rather an expectation value.

Returning to the statement that no physical law prevents MQS, it can be re-stated that for a coherent state (principal) in a closed system, the system evolves in time as the coherent state (plus an insignificant phase factor). More formally, the Hamiltonian or unitary evolution of the density operator for a closed system is given by:

$$\hat{\rho}(t) = U(t)\hat{\rho}(0)U(t)^{\dagger} = -\frac{i}{\hbar}[H, \hat{\rho}(t)] \qquad (3)$$

It should be clear that if the density operator and Hamiltonian commute, the expectation value of the density operator does not change in time (closed system or constant energy). This is independent of size and we have already illustrated how coherent systems can be expanded indefinitely without decoherence. But when the density operator of the principal is entangled with another system/environment (and therefore an open system with respect to the principal), the progression of the principal with time is traditionally represented by:

$$\hat{\rho}_S'(t) = tr_{env}[U(\hat{\rho}_S'(0) \otimes \hat{\rho}_{env})U^{\dagger}] \qquad (4)$$

Here, the environment is traced out at the given time point and the result is a new density operator 'representative' of the principal, which is no longer the same system. The 'real' principal cannot be described independent of the entangled environment so it is a representation (an expectation value). This representative principal progresses in time in a non-unitary manner while the joint density operators (a new single system) progresses in a unitary manner. If the entangling system is large and fluid relative to the principle, coherence is lost and this is the reductionist view of decoherence. But as alluded to in the interferometer study above and will be described under master equation approximations below, when the combined system forms a new closed system, a new larger coherent system develops which may represent an expansion of the principal. Therefore, in a complex system, it is unlikely the entangling system can be treated as simple and homogenous so that reductionist logic breakdown.

Therefore, in this simple model if there is no entanglement with the environment (or another system), the system remains coherent. But, if environmental entanglement occurs, the coherence may be lost as it is no longer describable independent of the entanglement. We will see that like the Young's interferometer example above, this density operator representation alone is still a reductionist approach to decoherence even though it allows for multi-particle interaction.

The focus of the paper is on the need for a complex systems or top-down approach to decoherence. Equations 3 and 4 together describe that a closed coherent state, of any size, remains coherent unless an environmental entanglement occurs. This is the reductionist model of decoherence that master equations are based on. If the environment is large, the model basically assumes the coherence dissipates into the environment. If the environment is limited, the coherence can be maintained in a new but larger form (or as we will see under compensation mechanisms, may correct coherent losses). We and other groups have discussed how in the absence of decoherence, a coherent system can be expanded to a macroscopic closed coherent system [10,14,19,22, and 23]. But as will be seen, even in this work the characteristics of a complex or CAS are not accounted for.

IIC. Decoherence Master Equations and Modeling

This section introduces master equations, which are essentially extensions of what were discussed in the previous section, but use approximations to be solvable making them nonlinear differential equations. Much of the paper deals with the lack of utility of master equations in complex systems, so their principles are reviewed here. Master equations are intended to directly yield the time evolution of the reduced total density operator (of the principal) for an open quantum system (interacting with an environment). Their assumed theoretical advantage is that knowing the dynamics of the total principal-environment is not needed as the environment is traced out. Even before going into specific in part III, it can be noted from what has already been discussed that ignoring the environment is not viable for a complex system (it ignores coherence preserving properties of the environment).

Equation 4 can be rewritten in a general form that does not explicitly include the environment. These master equations (in their most general form) use the assumption that the time evolution operator is time independent:

$$\hat{\rho}'(t) = J(t, t_0)\hat{\rho}'(t_0) \quad (5)$$

Here, J is a superoperator that dynamically maps the evolution of the principal. Obviously, based on equation 4, the superoperator J is generally impossible to solve without approximations. Therefore, the power of master equations comes from certain relatively simple assumptions about system-environment states and dynamics. But it will become equally obvious that as the complexity of the system increases, the validity of these approximations collapses. These assumptions give an approximate time evolution of the principal density operator for a reductionist system. The approximation equation is in the form of a first order differential equation artificially separating the coherent and decoherent aspects:

$$\frac{d}{dt}\hat{\rho}_S(t) = \underbrace{-i[\hat{H}'_S, \hat{\rho}_S(t)]}_{\text{unitary evolution}} + \underbrace{\hat{D}[\hat{\rho}_S(t)]}_{\text{decoherence}} \quad (6)$$

Here, ignoring the decoherence term, the remainder is a unitary evolution given by the commuter of an unperturbed Hamitonian and a Lamb-shifted density operator (the shift can generally be ignored). In other words, it is equivalent to equation 3. The second term is the decoherence term which is non-unitary. Among the most common approaches for making this manageable is to use the Born and Markov approximations (discussed in more detail below) [79,80]. Briefly, the Markov approximation corresponds to the assumption of a rapid decay of environmental self-correlation functions relative to the timescale set by the evolution of the principal (i.e. the environment is large compared to the size of the principal so information is dissipated). The Born approximations assume the system-environment coupling is weak or the density operator for the environment does not change significantly as a consequence of the interaction with the system. These assumptions were implicit in the Young's experiment above. We will discuss why the Born and Markov approximations are invalid for complex systems and CAS. In addition these master equations ignore emergent behavior, path integrals (with a few exceptions), compensation mechanisms, and the subentity problem.

We will see master equation predictions fail not only at the level of the quantum subsystems for photosynthesis and magnetoreceptor studies, but even simple Brownian motion. But with regard to the former, several papers on these quantum subsystems state that the efficacies measured are much greater than those estimated from master equations. It is common in these papers to get speculation of concepts like 'protein shielding', which is essentially a classical explanation [36,37,82]. This is reductionist thinking and likely the answer lies in complex decoherence compensations. Furthermore, these are significantly isolated systems implying even lower decoherence rates for intact systems. This supports the premise of the paper.

IID. Complex Adaptive Systems (CAS) and Emergent Behavior

This paper focuses on advancing MQS by moving away from reductionist analysis to focus on complex and CAS, both in biological and non-biological systems [47-59]. But in particular, as biological systems are almost uniformly CAS so we focus on them for initial insights into MQS research. This is because their sophisticated CAS, honed by evolution (natural selection), is more likely to support MQS and can be the source of design for non-biological systems. CAS research continues to expand across a wide range of fields. The principles span areas from condensed matter physics, the stock market, insect feeding, ecosystems, neurology, immune responses, and human social systems (ex: political organizations and traffic). CAS are capable of higher order behaviors such as emergence, robustness, compartmentalization, self-organization, and adaptability under extreme conditions. This is why reductionist modeling is not applicable. There is currently no uniform definitive definition of CAS, but one from John H. Holland from the Santa Fe Center should suffice for here [56]: "A CAS is a dynamic network of many agents (which may represent cells, species, molecules, individuals, firms, or nations) acting in parallel, constantly acting and reacting to what the other agents are doing. The control of a CAS tends to be highly dispersed and decentralized. If there is to be any classical coherent behavior in the system, it has to arise from competition and cooperation among the agents themselves. The overall behavior of the system is the result of a huge number of actions made every moment by many individual agents." It should be clear this is in complete contrast from the reductionist approaches described in the previous sections. What distinguishes a CAS from a pure complex collection or a nonadaptive multi-agent system (MAS) is the focus on top-level properties and features like emergence, adaptability, communication, compartmentalization, many autonomous parts, robustness, and self-organization. <u>Most of the phenomena dealt with in the human body owe their existence to CAS, and we expect decoherence compensation to be no exception</u>. It should be apparent that trying to understand how decoherence is dealt with in CAS, extrapolating decoherence theory from subatomic or simple systems (reductionist approach) as we saw in previous sections, is highly unlikely to yield useful results. It is analogous trying to predict the robust, self-organizing, and adaptive behavior of hurricanes from lab studies of water and air in beakers [54]. A top down approach is needed.

A common illustrative model of CAS involves bee function and it will be used here. Beehive temperature control is an example of a CAS exhibiting much of the higher-level behaviors relevant to this paper, albeit in a far simpler system [81]. For reproduction, the temperature in the hive needs to be tightly controlled (the reproductive capability here will be analogous to the coherence). If the temperature external to the hive begins rising, bees will start flapping their wings, which reduces temperature. The bees will be the ancillary (discussed in depth below), a part of the environment used to protect the coherence (the reproduction machinery). The key is heterogeneity in the threshold with which bees begin flapping their wings. Then as the temperature increases, the number of bees that are recruited increases. If the bees were homogeneous in response, when the temperature increased, the majority of bees would be recruited and there would be no fine control of temperature. This demonstrates robustness to external temperature and an emergent, adaptive behavior (temperature control) by variable response of the ancillary. It is not predictable by analyzing individual bees, similar to the flaw in extrapolating data from atomic systems to much larger systems.

Similarly, there can be predators, which can be viewed as analogous to environmental entanglements causing decoherence. Almost no single bee is critical to the hive function in response to predators. So if predators remove a number of the bees (ancillary) less than a certain threshold, the reproductive capacity is preserved by redistributing the remaining bees. In addition, when a predator approaches a hive, pheromones are released which concentrate the bees capable of jointly stinging the predator. The reproductive machinery, the analogy to coherence, is protected by mechanism not predicted by reductionist studies (examining the properties of individual bees). These are emergent behaviors only seen in complex systems. We will summarize this simple CAS example relative to decoherence. First, temperature control and defense are emergent behaviors. If we had studied individual bees and tried to extend the results to the whole system behavior (reductionism), it is highly unlikely we could have predicted these higher order behaviors. Second, there is robustness to reproduction as the hive (ancillary) is able to maintain via emergent behavior in the setting of ancillary/bee loss to predators (environmental entanglements). The emergent behaviors in the bee colony, such as temperature control and predator response, would not be apparent from the study of a few elements (reductionism) and extrapolating them to a larger scale.

III. Failure of Reductionist Approaches and Their Master Equations for Complex Systems

**With the overview in the two previous introductory sections, we move to the focus of the paper. This is the failure of reductionist decoherence approaches, and their master equations, with complex systems and the need for more top-down studies.** In previous paragraphs, overviews were given of coherence/decoherence from a reductionist standpoint, the concept of master equations, and the basic principles of a CAS. Through the remainder of the paper, the inadequacy of reductionist approaches to deal with properties of CAS (emergence, adaptability, communication, compartmentalization, many autonomous parts, robustness, and self-organization) is discussed. Furthermore, in the next several sections, we will discuss how decoherence master equations fail for CAS <u>beyond</u> not accounting for emergent behaviors. This includes the inadequacy of the commonly used approximations with CAS, not taking into account decoherence compensating mechanisms, not generally accounting for path integrals, and the subentity problem. **We propose accounting for these factors with a top-down approach, modeled by finding equivalent classes and transition functions that result in the development of homomorphisms.**

a. Examples of Reduction Approach Failures in Classical Systems

The focus of the discussion will be the inapplicability of current decoherence theory and decoherence master equations for modeling (as well as experimental design) with complex and CAS. Before discussing quantum systems, we will discuss some of the many reductionist failures in classical mechanics. <u>When a scientist faces a complex world, traditional tools that rely on reducing the system to its fundamental elements generally allows them to gain insight.</u> This of course is, has, and will be a critical approach to a substantial portion of scientific inquires. <u>Unfortunately, using these same tools to understand complex systems (particularly CAS) fails, because it becomes impossible to reduce the system without killing many of its essential features.</u> This sentiment that macroscopic systems can't realistically be calculated up from the physics of atomic and subatomic particles has been expressed by, among others, two Nobel Laureates in particle physics, PW Anderson and M. Gell-Mann. They used the example that chirality which can't be predicted from subatomic particles. A top-down approach is generally against the standard reductionism one, where the latter tries to decompose any system to its constituents and hopes that by understanding the elements or subsystems, one can understand the whole system. This line of reasoning fails for many classical situations, three of which will be listed here for analogy. In the first example, it could be stated from a reductionist approach that since polyunsaturated lipids (ex; arachidonic acid) rapidly breakdown when exposed to general environments, such as exposure to one oxygen molecule, they cannot be important biomolecules. Yet they remain stable within the lipid environment of biological membranes indefinitely, among other locations, and are important precursors of prostaglandins and thromboxane. Biological systems compartmentalize these molecules through non-trivial approaches, which leads to robustness (this ability of biological systems to compartmentalize function will be discussed).

A second example of how the reductionist approach fails, where the whole system needs to be accounted for, is a recent study in *Cell* [47]. This article looked at how the local environment affects the DNA profile, rather than the reverse (DNA distribution generates the profile). They performed reciprocal transplantations of gut flora into germ-free Zebrafish and mouse recipients. The relative abundance of the lineages changes to resemble the normal gut microbial community composition of the recipient host rather than the original community it was transplanted from. Thus, differences in community structure between Zebrafish and mice arise in part from distinct selective pressures imposed within the CAS of the gut habitat of each host rather than the DNA profile introduced. A reductionist approach, such as growing the isolated flora in a Petri dish and extrapolating it to an in vivo distribution, would have predicted the opposite outcome.

Finally, with a non-biological example, hurricanes are robust, self-organizing, and adaptive, but we would not attempt to predict their behavior from extrapolating data solely from water and air in laboratory beakers. Hurricanes are studied from a top down approach, modeling with equivalent classes such as air pressure and water temperature [54]. All three examples are modeled by finding equivalent classes and transition functions that result in the development of homomorphisms. Expertise with this type of modeling is well established in the CAS field, and should be leveraged for MQS and complex decoherence work. Any complex decoherence theory needs to account for emergent (high-order) behaviors.

b. Approximation Failures

Even with the brief discussion of CAS presented here, beyond not accounting for higher order behaviors, the limitations of Markovian and Bornian approximations in trying to derive master equations for CAS are seen [79,80]. As stated, the Markovian approximation corresponds to the assumption of a rapid decay of the environmental self-correlation functions because the environment is large compared to the size of the system (essentially losing coherence into an infinite environment). This is generally not valid for a CAS where coordinating actions within limited areas is characteristic, for example with compartmentalization that is central to biological systems. If we look at microtubules (discussed below in more detail), which have been suggested in sustaining coherent behavior, they are as long as 25 mm but have an inner diameter less than 20 nm, with separations between tubules on the order of 20 nm. Unlike isolated atomic systems, it is difficult to justify the Markovian approximations under these circumstances. The Born approximations assume the system-environment coupling is weak or, put another way, the density operator for the environment does not change significantly as a consequence of the interaction with the system. This would represent a conflict with the definition of a CAS, where "The control of a CAS tends to be highly dispersed and decentralized. If there is to be any classical coherent behavior in the system, it has to arise from competition and cooperation among the agents themselves [56]." Extensive interaction and change within the 'environment' is a central component to CAS.

c. Not Taking into Account Decoherence Compensating Mechanisms

In addition to not taking into account higher order CAS behaviors and approximation failures, master equations also do not address potential mechanisms for decoherence compensation. Primarily from the field of quantum information systems, non-classically based approaches have been advanced for decoherence compensation [79,80,84-98]. It not unreasonable to consider the possibility biological systems with their CAS can utilize these or more likely more advanced approaches. Several of these approaches will be discussed here. Keeping consistent with the information systems literature, these approaches will be discussed in terms of information flow rather than the specific mediators carrying the information (the specific structure may not be constant). So, the progression or alteration of a component of information (analogous to a qubit), with time, will be used rather than mediators (like proteins). For those unfamiliar, the information in DNA that is translated through RNA to proteins would be considered an example of expressing the same information (essentially) flowing through a system in different forms. Another analogous example would be a song recorded on tape, then transferred over the internet (in another form), and ultimately being in MP3 format. The information is essentially the same, just changing in form. Both in the case of DNA and the song, as the information is changing embodiments errors (analogous to decoherence) may be compensated for or prevented. So it is easier to describe these decoherence compensation mechanisms in terms of information flow rather than structures, as done for quantum information systems, avoiding the need to specify specific mediators for now.

Before discussing decoherence compensation mechanisms, and the fact they are not incorporated into decoherence master equations, it should be noted there are certain prohibitive rules that exist with quantum information systems that have no classical analog (thereby preventing some forms of classical corrections). These rules include: 1.) A no-cloning rule that stems from the uncertainty principle. It prevents making repetitive exact copies of an original signal (i.e. creating direct code redundancy such as classical majority coding or majority voting). 2.) That a measurement almost always leads to loss of information, particularly quantum coherences. By its very nature, measurement eliminates the off diagonal elements of the density operator and selects a single eigenvalue as an output. This process destroys a substantial amount of information from the principal, primarily the quantum characteristics. 3.) That the errors are continuous rather than discrete.

So, while a classical bit-type error can only have values of 0 or 1, qubit errors can have these values or any superposition in between.

But as we will see, quantum mechanics offers its own unique laws for dealing with decoherence not available in classical mechanics. These will be used in the next several examples of decoherence compensation from the quantum communication literature. In particular, as can be seen, certain quantum laws make active interaction with the environment an important part of these compensation techniques. This is in contrast to reductionist approaches where the environment is almost always a source of information loss (which was preliminarily touched on above). We will define three components of the system, the information (coherent portion) known as the principal, the environment, and the ancillary. The ancillary is the portion of the environment that is used or becomes part of the principal. These physical elements can change from one role (ex: environment) to another (ex: ancillary) [79,80]. Example decoherence compensating methods are:

*1. Fault Tolerant Code.* A first principle in maintaining the principal, which comes from indistinguishability, is to make the code fault tolerant. In other words, the code can be made of a form where, even though interaction with the environment leaves it changed, the paths remain indistinguishable and therefore coherence information maintained. This is analogous to what was seen with the Young's experiment above or Brune experiments referenced as examples (8, 15, 16, 22, 23, 136).

*2. Redundancy Without Cloning:* In telecommunication, the diversity of information that needs to be transmitted is generally vast. This is not likely the case for biological system because most functions can be accomplished classically (which seems to be the case for photosynthesis and magnetoreception). So for a given subsystem such as the leukemic cell cultures to be discussed, it is reasonable to postulate that the number of commands needing transmission through quantum systems is relatively small in number (multiply, apoptosis, etc). This is also not an unreasonable postulate for CAS as often a small number of interactions control the complex behavior, such as the small number of commands/interactions that drive the beehive in the example above [81]. Therefore, for a classical system, this would allow for the use of codes of relatively large size with substantial redundancy such as 1111111111, 0000000000, 1111100000, and 0000011111, repeated at a high frequency. Then if errors are in a single bit, they are easier to identify and correct in this form. However, the no cloning rule prevents this from being done directly. Instead, the initial qubit (and this can be extended to multiple qubits) is combined through a unitary operation with a specific number of ancillaries (again, portions of the environment are now being used as part of the principal) producing a redundant code. The code can then be even further concatenated to improve the fault (decoherence) tolerance in terms of size and redundancy. With this approach, the final code is distinct from the initial code in form, but it encodes the same information in a highly redundant manner. This form of decoherence compensation is not accounted for by the master equations.

*3. Dealing with Continuous Error.* Qubit errors (ex: spin states), as stated, are continuous and can be a bit flip, phase flip, or a combination (not counting classical errors). However, due to the unique properties of quantum mechanics (such as the relationships among spins), these quantum errors can be corrected through the process of discretization of errors. The process uses Kraus operators (the operators, as well as how they are used, are described elsewhere) and takes advantages of the Pauli spin operators' properties. This approach allows errors of any variation or degree (such as a three degree phase shift) to be corrected. How this approach is executed requires considerable explanation and is discussed elsewhere, but it demonstrates a correction technique unique to quantum mechanics with no classical corollary [79,80].

*4. Correction without Measurement.* As stated, measurement destroys information; so making a measurement to correct errors would not be useful for maintaining coherence within complex systems. However, there are techniques using ancillaries that allow the errors to be transmitted to the ancillaries, and principal information is restored, thereby avoiding a principal measurement that would break it down into a classical state. The ancillary's high degrees of freedom means errors of the principal are lost into the environment, retaining path indistinguishability. Again, this is not accounted for in modeling by master equations.

*5. Decoherence Free Spaces.* Another possibility for dealing with coherence loss is decoherence-free subspaces that are covered in more detail below. In the effort to build a quantum computer, much attention has been given to identifying subspaces of Hilbert space that are unaffected by the coupling of the system to its environment. For example, paradoxically, when a system couples very strongly to its environment through certain degrees of freedom, it can effectively "freeze" other degrees of freedom by a sort of quantum Zeno effect, enabling coherent superpositions and even entanglement/coherence to persist (86-97).

A clear example is provided by a double-well one-dimensional potential (98,99). A system placed in the lowest energy state of one well will tunnel back and forth through the intervening barrier, oscillating with a certain frequency. If the particle is placed instead in an excited state of the well, this flip-flop frequency will be different. Thus, an initial state consisting of a superposition of lowest energy and excited states will soon evolve into a complicated muddle as the flip-flops get out of phase. *However, if the particle is allowed to interact strongly with the environment, this interaction has the effect of forcing the disparate oscillations into synchrony, thereby maintaining a limited form of quantum coherence, not only in spite of, but because of, environmental interactions.* Furthermore, if the system is placed in quantum-correlated state of left and right wells, this 'less stable' correlation is also preserved by environmental interaction. The model was developed in the context of neutrino oscillations, but has general applicability (98,99). Decoherence free subspaces are further discussed below.

6. *Summary*: Man-made techniques for dealing with decoherence have been developed, primarily in the quantum communication field, but are generally not considered with reductionist theory.  It is likely that evolution has led to even more sophisticated approaches for biological systems to deal with decoherence in CAS.  As with other emergent behaviors in CAS, these approaches would likely be inconceivable from a bottom up approach.  A top down analysis is needed to understand systems where these and other types of large-scale compensatory mechanisms exist.

d. Master Equations and the Use of Path Integrals

In the last several paragraphs we discussed why reductionist approaches (extrapolating data from atomic and subatomic systems to complex systems) and their associated master equations do not generally take into account emergent properties from CAS, approximation failures when dealing with complex systems, and already known decoherence compensation approaches.  This lack of effectiveness is particularly relevant to establishing experimental designs for identifying or supporting MQS.  Later in this paper we will produce experimental data illustrating how a top down approach can look for quantum non-locality in a biological system (part IV).  In this section, we address another limitation of reductionist approaches to decoherence in complex systems.  This is the ignoring of path integrals (or more precisely membrane integrals) in most master equations, particularly path integrals that take into account perturbations (with decoherence being the perturbation here).  In the Young's example above environmental interactions needed to be accounted for in two paths.  In reality it needs to be accounted for in all potential paths.

Master equations in the form of canonical variables are generally a particle representation, but in CAS a field representation is required (as can be seen from the quantum mesoscopic exciton systems of photosynthesis).  One important method for dealing with decoherence may be maintaining a modified action.  Specifically, in the Young's example above, the two slits reduce the number of paths (in the field integral) but a coherent action is maintained.  Decoherence may operate in an analogous manner to the slits (potential paths) which would not be seen in reductionist systems.  A coherent action is maintained in spite of some path loss from decoherent interactions.  The action can ultimately be maintained from this path integral prospective if we view decoherence as disrupting, but not completely destroying, all paths.

The path integral [100-104], as well as the non-local nature of quantum mechanics [1-19, 105-106], has arisen as fundamental in quantum field theory [100-107].  Some have suggested that path integrals are even more fundamental than space and time, so their role in decoherence is discussed here in some detail [101].  But the key point, it should become clear that when looking at complex systems with environmental entanglements (particularly CAS), determining the action from the path integral is likely a futile effort (reductionist failure), supporting a top-down approach for understanding the system.

The path integral approach has been applied only minimally to decoherence master equations, with the description of quantum Brownian motion being perhaps the most significant example [103-104].  But even in this relatively simple system, it is far from completely described (discussed below).  The decoherence is expressed in terms of the influence functionals.  This can be thought of as a dynamic sculpturing of the action and coherence.  Of note is that a path integral representation can be converted into canonical variables or Schrödinger's representation, but the reverse is not necessarily true. Specifically, the Schrödinger equation is a diffusion equation with an imaginary diffusion constant, and the path integral is an analytic continuation of a method for summing up all possible random walks.  Schrödinger's equation is also not valid under relativistic conditions, emphasizing another way the path integral is more fundamental.

So reviewing, the path integral formulation of quantum mechanics (begun by Dirac but brought to a practical form by Feynman) is a description from quantum theory that generalizes the action principle of classical mechanics [100-101]. It replaces the classical notion of a single unique trajectory for a system with a sum, or functional integral, over an infinite number of possible trajectories in order to compute the quantum amplitude. The basic concepts will only be touched on here, but can be found in many other sources in more detail. Feynman proposed to recover all of quantum mechanics from the action using the following postulates:
1. The probability for an event is given by the squared length of a complex number called the "probability amplitude".
2. The probability amplitude is given by adding together the contributions of all the histories in configuration space.

3. The contribution of a history to the amplitude is proportional to $e^{iS/\hbar}$, where $\hbar$ is reduced Planck's constant (whose units are the same as the action), while S is the action of that history, given by the time integral of the Lagrangian along the corresponding path.

In order to find the overall probability amplitude for a given process, then, one adds up, or integrates, the amplitude of postulate 3 over the space of all possible histories of the system in between the initial and final states, including histories that are absurd by classical standards. In calculating the amplitude for a single particle to go from one place to another in a given time, it would be correct to include histories in which the particle describes elaborate curlicues, histories in which the particle shoots off into outer space and flies back again, and so forth. The path integral assigns all of these histories amplitudes to equal magnitude but with varying phase, or argument of the complex number. The contributions that are wildly different from the classical history are suppressed only by the interference of similar, canceling histories. Feynman showed that this formulation of quantum mechanics is equivalent to the canonical approach to quantum mechanics, when the Hamiltonian is quadratic in the momentum.

The path integral for a closed system can be represented by:

$$\psi_t(y) = \int \psi_0(x) \int_{x(0)=x}^{x(t)=y} e^{iS} Dx \qquad (7)$$

The wavefunctions are the initial and final states and $D_x$ is the integral over all paths. As stated, the path integral has been applied to Brownian motion, including attempting to account for decoherence, which can be considered an open system that is brought to a closed function by the path integral. The effect of the external quantum systems are <u>approximated</u> with a class of functionals, the influence functionals, of the coordinates of the system only, and are used here to represent decoherence.

To address decoherence through path integrals, we again begin with the complete (unsolvable) master equation (which is equation 5):

$$\hat{\rho}(t) = J(t, t_0)\hat{\rho}(t_0) \qquad (8)$$

The time evolution operator (in path integral form) is given by:

$$J\langle x, q, x', q', t | x_i, q_i, x_i', q_i'.0 \rangle = \int_{x_i}^{x} Dx \int_{q_i}^{q} Dq \exp\left[\frac{i}{\hbar} S[x, q]\right] \int_{x_i'}^{x'} Dx' \int_{q_i'}^{q'} Dq' \exp\left[-\frac{i}{\hbar} S[x', q']\right]$$

(9)

All symbols have been previously described, where this is an extension of equation 7. However, this form does not contain a decoherence term, which we will introduce (influence functionals). The first approximation generally made for decoherence equations, with respect to path integrals, is that the action of the principal, environment, and combined interaction can be treated separately:

$$S[x,q] = S[x] + S_E[q] + S_{int}[x,q] \qquad (10)$$

Now, the evolution operator is unsolvable in the form of equation 9 so previous work has used another approximation, a reduced evolution operator as well as a reduced density operator. The limitations of these approximations are discussed in the next section. The reduced evolution operator, attempting to account for environmental interactions and with the above approximations becomes:

$$J_r \langle x_f, x_f', t | x_i, x_i', 0 \rangle = \int_{x_i}^{x_f} Dx \int_{x'_i}^{x'_f} Dx' \exp\left[\frac{i}{\hbar}(S[x] - S[x'])\right] F[x,x']$$

$$= \int_{x_i}^{x_f} Dx \int_{x'_i}^{x'_f} Dx' \exp\left[\frac{i}{\hbar} A[x,x']\right] \qquad (11)$$

Where $A[x,x']$ is the effective action for an open system and $F[x,x']$ is called the influence functional, which is being used here primarily to represent the influence of decoherence. The functional, using the separable action of equation 10, is given by:

$$F[x,x'] = \int_{-\infty}^{\infty} dq_f \int_{-\infty}^{\infty} dq_i \int_{-\infty}^{\infty} dq_i' \int_{q_i}^{q_f} Dq \int_{q_i'}^{q_f} Dq' \exp\left[\frac{i}{\hbar}(S_b[q] + S_{int}[x,q] - S_b[q'] - S_{int}[x',q'])\right] \rho_b(q_i, q_i', 0) \qquad (12)$$

This is a field representation that can be generalized to a system interacting with its environment. Several problems are already clear:

1. The master equation, though more accurate, is not completely solvable even for a simple system like Brownian motion, even taking at least partially into account path integrals (using the approximations of equation 10-12). Therefore, even though there is a need to account for path integrals in master equations, how can the reductionist results be expected to extend coherent behavior (and design experimental set-ups) in CAS, if unsolvable? The system needs to be modeled top down. The influence of path integrals in the system, as with other limitations of master equations addressed, needs to be incorporated in homomorphisms for top down analysis.
2. Reduced density and time evolution operators are used, the limitations of which are discussed in the next section.
3. The equation makes the assumption of a linearly separable action as per equations 10 and 12, which does not hold for a complex system (by the definition above these systems are highly interactive).
4. In our example of Young's experiment, we saw that decoherence results in a simple system when the inner product is zero between the system and environment. This degree of interaction with the environment is not easily addressed in equation 12, as in this form it is essentially all or none. This will be addressed again with another Young's experiment example below.
5. As others and we have pointed out in previous publications, environmental entanglements can reduce, expand, or have no effect on coherence. In two papers from our group in particular, if the entangling elements in two (or more) paths are identical from the prospective of the ultimate measurement (indistinguishable paths), in the setting of first order coherence (single particle wavepacket) and failure of common approximations (Born and Markovian), coherence is expanded rather than lost [22,23]. This is not accounted for in equation 12.
6. Decoherence compensating mechanisms are not taken into account.
7. It is assumed that at time zero, no correlations exist between the principal and the environment. If they do, decoherence may have no effect.
8. If the environment has certain properties, such as being supraohmic, decoherence may not occur from equation 12.
9. Markovian and Born approximations are still assumed, the problems with this have already been discussed.

These points illustrate yet another reason why the behavior of complex systems can not be adequately described by extrapolating data from the atomic and subatomic level to complex systems. The action can potentially be preserved (in complex systems) through distinct mechanisms that involve maintaining sufficient potential paths in the setting of an interactive environment. We previously used Young's experiment to illustrate loss of coherence with loss of path distinguishability via environmental entanglements (then the path integral was ignored). Similarly, we will now use it to illustrate qualitatively maintaining the action in the setting of environmental entanglements. This is seen in figure 2. In figure 2A, Young's experiment is depicted, except with a partial path integral (12 paths rather than an infinite number for illustration) rather than a single path. Note that the path integral for the Young's experiment is described from the source to the screen involving both slits (i.e. the action involves both slits). There are no decoherence/environmental entanglements in figure 2A, so an interference pattern exists.

Now in figure 2B, the action and interference pattern are maintained in spite of a reduction of some paths from decoherence. In figure 2C, coherence and interference are present, but altered, as the action has been phase-shifted (coherence in a new form). Therefore, destruction of some paths does not necessarily correspond to loss of coherence or the action. We can view this as the inner product being not being zero for all paths, or that some cover decoherence free subspaces. Again, decoherence free (or reduced) subspaces are subspaces of Hilbert space of the system in which every state in the subspace is immune (resistance) to decoherence. Examples were listed above. Formally, for these subspaces the preferred states of the principle are defined by (the general diagonal decomposition):

$$\hat{S}_\alpha |s_i\rangle = \lambda^{(\alpha)} |s_i\rangle \text{ for all } \alpha \text{ and } I \qquad (13)$$

The eigenvectors form an orthogonal basis of the subspace and the eigenvalues ($\lambda$) are independent of the index (simultaneously degenerate, no i subscript). It can be interpreted that there is no term in the interaction Hamiltonian that would jointly act on both the principal and the environment in a nontrivial manner. Or alternatively, if we consider the principle a collection of qubits, the qubits are considered extremely close relative to the coherence length of the environment.

So, three central points are important for path integral-based with respect to decoherence master equations limitations for complex system (and the study macroscopic quantum phenomena). First, accounting for path integrals is needed but rarely done. Second, even when path integral master equations are used, they still cannot give a complete description of a CAS because they are unsolvable, therefore they can not provide sufficient insight into the ability to sustain macroscopic quantum phenomena (or for experimental design). Third, common master equations such as those not incorporating the path integral do not allow for this action 'tailoring' by decoherence, where some, but not all, of the potential paths are removed with environmental interactions (a situation where the action and coherence are maintained).

e. Use of Reduced Density and Evolution Operators and the Subentity Problem

The inability of reductionist approaches to effectively deal with CAS has been discussed in terms of not accounting for emergent behavior, inappropriate approximations, compensatory mechanisms, or path integrals. The final point on reductionist limitations is more challenging to appreciate (abstract), but it emphasizes an additional challenge in addressing MQS in CAS from a reductionist approach.

This section will deal with two related but distinct issues surrounding the nature of a quantum state without measurement, which will be discussed in terms of the density operators. This is the subentity problem that has at least two components. 1.) The use of reduced density and evolution operators in master equations. 2.) The meaning of a density operator of a mixture without measurement. While they both technically fall under the category of the subentity problem, many authors limit this to the latter. It basically comes down to an inability to deal with quantum states in classical environments that maintain their "quantumness".

Recall four points already discussed with respect to density operators. First, that the density operator of quantum mechanics is not equivalent to the density operator of statistical mechanics (the former is an operator on Hilbert space). Its off-diagonal elements are not real. Second, the trace operation of a density operator/observable product gives an ensemble average (equation 2). Third, the trace operation is frequently used with entangled states (such as in decoherence) or other inseparable density operator states to give a representation of a subsystem (equation 4). However, this trace representation is a hypothetical ensemble average and not a 'real' state (a mathematical tool). The fourth point is they are almost universally used in decoherence master equations, relying on reduced density and time evolution operators, but are an approximation and not an actual representation of the state (reduced operators utility becomes even less as the system becomes more complex). This was seen when examining quantum Brownian motion (equation 12).

The fact that the output of the trace operation produces an ensemble average and not an actual value of the principal (or evolutionary operator) has already been addressed. Non-local entanglement progresses forward in a unitary manner, but the reduced density operator does not, illustrating the weakness of the approach of using the trace to represent subsystems (does not represent the full quantum entity). The reduced density and time evolution operators have actually lost coherences and indistinguishable paths in this approximation. There is no way to represent a subsystem of an entanglement (including decoherence) accurately, it loses the quantum mechanical aspects when represented by an ensemble average. So in the central equations for reductionist decoherence, the 'quantumness' has been removed in order to make the equations more manageable and give real values. This approximation may be useful with simple systems, but with complex systems this approximation of quantum states as classical states on a large scale is unlikely to produce meaningful modeling/understanding and experimental design.

This point was expanded on by D'Espagnat (as well as Hughes) with respect to *mixtures* [77-78, 108-111]. They discounted von Neumann's ignorance interpretation (of quantum mixtures) of the trace operation results [77-78]. D'Espagnat believed all mixtures are improper without measurement (they are pure states) and cited the fact that you cannot reconstruct the total system from the subsystems of the trace operation (essentially the same subsystem problem). This re-enforces the view that it is unlikely these reductionist approaches (using reduced density operators) can be extended to predict the handling of decoherence in CAS because they do not represent the 'true' principal. The coherence, even of a mixture, would progress in time as pure state with imaginary components until the equivalent of measurement is performed. Again, this supports the point that MQS in CAS needs to be evaluated from a top-down approach, as a limited reductionist view of the world cannot be extended up to complex systems to predict the ability to maintain coherence.

**Part IV: Experimental Logic and Results**

The previous section focused on the failure of reductionist approaches and their master equations for addressing decoherence in CAS. In this section we present a top down experiment as a model for looking for MQS in CAS. In addition, we will give an overview of the top-down or totalistic experiment.

IVA. Top Down or Totalistic Approach

For the reasons the authors already listed alone, stating a quantum phenomena is not capable of existing at a macroscopic scale because of wet environment is not justifiable based on extrapolating reductionist decoherence (or calculations from master equations). It is far more likely that they are not easily detectable in the context of a large intact classical system or destroyed by partially isolating the system.

The goal with a top down approach, in contrast to the reductionist approach, is to use information observed from analysis of intact or relatively intact systems in response to differing conditions/agents. The quantum mesoscopic studies of photosynthesis and magnetoreception, described in the appendix, are examples. A database of responses to stimuli, at the tissue or organism level, is accumulated with the top down approach. The information obtained is used to establish refined future experimental designs as well as deeper insight. After extensively assessing the data from this totalistic approach, which requires large numbers of experiments, the information is used to begin defining more and more mechanisms. It is then that reductionist approaches have a role as seen in mesoscopic quantum systems (ex. photosynthesis and magnetoreceptors) or in classical systems (ex. Mendelian inheritance).

IVB. Experimental Design, Data Analysis, and Modeling

When you approach CAS, the experimental design is that top down testing generally begins with educated postulates, which we will see in the representative study in this paper. In general then, the experimental designs become more refined to maximize outcomes based on the previous results. A classic example of the success of this approach is Mendelian inheritance, which began with hybridization experiments in garden peas (published in 1866). Observations led to experimental designs that were gradually refined, more clearly defining mechanisms of inheritance. After many decades with a multitude of top-down experiments, DNA was ultimately characterized with a reductionist studies tailored to the top-down results. Knockout mice and the Hamster Cheek pouch models are additional examples of top down approaches, looking at responses in nearly intact systems varied just enough from 'normal' to allow parameters of interest to be observed within the complex system. Perhaps more relevant, because they deal with quantum mechanics, are the over 100 years of top down research studies that ultimately led to the identification of mesoscopic quantum systems in photosynthesis and magnetoreception. Reductionist studies eventually performed were tailored based on the multitude of top down studies over many decades. In the experimental approach used in this paper, intact leukemic cells were used to demonstrate non-local quantum correlations in cell mortality where no contact or exchange of biomolecules occurred. We focused on the optimizing conditions in these top-down studies, making educated initial postulates on parameter values such as field exposure and percent confluence, to increase the likelihood of a successful experiment (non-local correlations in cell mortality).

The goal in early modeling of complex systems is finding equivalent classes and transition functions that result in homomorphisms (and not trying to scale up from subatomic particles). A homomorphism is a transformation of one set into another that preserves within the second set the operations between the members of the first set. An example would be using momentum exchange or wind-sea enthalpy in predicting hurricane movement as opposed to trying to make predictions based the properties of elementary water and air molecules. Examples of the modeling approaches to these homomorphisms include networks, cellular automata, multi-agent system, and computational biology. Expertise with this type of modeling is well established in the classical CAS field, and should be leveraged for MQS and complex decoherence work. This would allow experimental conditions to be optimized and predicable patterns to be elucidated without destroying the CAS, eventually narrowing down to specific mediators with ultimately tailored reductionist studies.

IVC. Challenges of Macroscopic Quantum Biology and other CAS

The reasons for using biological systems to look for MQS are discussed above. But it is not surprising that if the hypothesis is correct and macroscopic quantum phenomenon do occur in biological systems, their reproducible identification would be challenging. The reasons include the following. First, because biological systems are complex, characterized by emergent behavior and robustness, substantially reductionist designs are not possible to identify mechanisms. Second, a small amount of system isolation is needed to identify mechanisms, but this isolation needs to be carefully designed. This is because the system may become vulnerable to small external influences (such as electromagnetic fields) making initial success and/or reproducibility a challenge. It is assumed that if classical molecules like polyunsaturated fats are unstable in air or water, vulnerability is an even worse problem for quantum phenomena. Third, even if a phenomenon is identified, it is difficult to differentiate confidently if it is classical or quantum mechanical in an intact system.

The previous paragraph pointed out the challenges of identifying or sustaining MQS in complex systems. But it is the prevailing opinion that controlling MQS would represent a paradigm shift across many disciplines, including biology. Among the areas where macroscopic quantum systems show the most promise are delocalized signaling, cell/tissue energy efficacy, and non-local remote communication [60,112]. In this paper our group will present data from a small study consistent with non-local quantum correlations between remote leukemic cell (HL-60) populations resulting in correlated mortality. Though the results are not definitive, needing further studies, they demonstrate the utility of the top down approach for studying potential biological quantum systems.

IVD. Preliminary Summary of Results

To summarize the results of the experiment to be discussed below, cell death was induced in one population (leukemic cell population) by a microtubule directed apoptotic agent (Taxol) [118]. This also led to death in a spatially separated (remote) cell population with no direct contact or exchange of biomolecules. This cell death did not occur in controls. We postulate the results are due to quantum correlations between mesoscopic subsystems in the different populations. We are also postulating the most likely mediators are cryptochrome within the classical cell population environment (but have not excluded other subsystems such as microtubules). But as will also be seen, while the experiments are consistent with quantum correlations, intrinsic low frequency electromagnetic fields (LFEF) have not been excluded so further studies will be needed. But this is consistent with top down studies with the experiment moving the field forward in steps to the underlying mechanism.

IVE. Rationale for the Experiment:

We are using this as an example of attacking a potential quantum system through a top-down approach. An important question in moving forward with our experimental model/protocol is why would we postulate white blood cells (WBC) and leukemic cells exhibit non-local quantum correlations? The reasons include:

1. *Why HL-60 cells?* We have anecdotal observations spanning several decades that HL-60 cells exhibit non-classical correlations in cell mortality.
2. *Is there a reason why nature would provide neutrophils with non-local signaling?* Yes, neutrophils are spread throughout the body and generally with no contact. Non-local signaling would reduce collateral tissue damage when inflammation is scaling down (apoptosis), coordinate cell entry into the circulation (or a region), coordinate their peripheral clocks independent of the central circadian clock, and control proliferation.
3. *Do quantum correlations, the two potential mechanisms being examined, have an advantage over traditional signaling methods*? Yes, quantum correlations are essentially instantaneous, long range, and would not be altered by traditional barriers (such as the blood brain barrier or basement membrane). But it is assumed most communications is classical.
4. *What are the potential mediators?* As stated, we strongly favor cryptochrome but are considering two other mediators, the $\pi$ cloud of microtubules and membrane vibrational modes.
5. *Why focus on cryptochrome*? Cryptochrome is the primary mediator of most magnetoreception in biological systems. First, the spin state distribution and spin precession are both extremely sensitive to LFEF (at different frequencies). Second, the spin states are capable of non-local quantum spin correlations, which we postulate is the most likely signaling mechanism. Third, cryptochrome is a central mediator in the cell cycle, local circadian clock, master circadian clocks, and apoptosis. So we will use cryptochrome's properties to control cell function remotely (extrinsically).
6. *Why has this signaling not been identified before?* We are unaware of any studies, with positive or negative results, examining non-local, biomolecule independent signaling. This is likely for similar reasons to why the magnetoreception field progressed slowly; the mediators/mechanisms will not survive reductionist approaches and are masked by the larger classical systems (requiring a top down approach).
7. Primarily from optical experiments performed for well over a decade, non-local correlations have been demonstrated between remote objects (non-biological), such as mirrors, though under far from ambient conditions. This includes work from our group. [1-21, 22, 23, 107, 135-137].
8. Non-locality is fundamental to quantum mechanics. Concepts of local realism are still discussed in the scientific community, but are becoming more the prevue of internet discussions rather than peer-reviewed journals. The experimental evidence supporting non-local quantum correlations now stretches several decades. Work with macroscopic systems under ambient conditions lends confirmation, including publications particularly relevant to this paper by our group with remote reflectors, a polarized light study, and a recent study in *Science* with diamonds [75, 76, 135-137]. It should be noted that in these three studies, when correlations were established between remote objects, actions on one led to observations of changes on the other remote system, similar to the study in this paper [135-137]. This is in addition to work under far from ambient conditions [1-9, 17, 107]. Yet, for reasons that appear to have no solid scientific basis, reluctance exists in conceiving of non-local mesoscopic systems distinct from concerns over decoherence. This emphasizes the importance of work like that being presented in this paper, pointing out the lack of a solid scientific basis for resistance to what represents fundamental scientific principles.
9. There are various mechanisms for generating non-local quantum correlations, which we have recently reviewed, as well as have been identified by other groups, [10, 14, 19, 22, 23] that would be applicable to MQS.
10. Decoherence theory suggesting the universal failure of non-local mesoscopic/macroscopic quantum correlations (under ambient conditions) is reductionist, while a complex systems approach suggests prolonged coherence is potentially achievable. In other words, even for those who accept microscopic non-locality, discounting MQS non-locality is based on reduction decoherence views that predict rapid breakdown on larger scales.

IVF. Postulate in Design of the Top Down Study:

***The hypothesis addressed in the experimental portion of the paper is that one cell population can be altered in function by intervention on a second cell population non-locally without direct contact or biomolecule exchange.*** It is postulated that if the hypothesis is tested correctly the non-local interaction is quantum correlations between subsystems such as cryptochrome. It is consistent with the complex decoherence focus of the paper.

Quantum non-locality has not been published in biological systems, so several educated postulates based on theoretical and other experimental considerations were made (and are essential). These are examples of the type of postulates needed in top-down experiments (along of anecdotal observations). *One postulate* was it would be more likely to identify a robust quantum effect in a partially isolated leukemic cell line than a solid tumor culture. The rationale is that coherence in leukemic cells would likely require less supportive structure. It was therefore postulated that leukemic cells, which are blood-borne, are not likely to need the same level of environmental support. *Postulate two* was that agitating the cells, to increase initial proximity, would improve the chances of establishing initial correlations (before the final split), whether by direct contact or mediators. *A third postulate* was that splitting cells in culture, without the high intensity field, maintained quantum correlations [63-66]. *A fourth postulate* was that exposure to a high intensity, LFEF would disrupt correlations between spatially separated populations [44]. The postulate is supported, for among other reasons, on a similar approach in disrupting magnetoreceptors (cryptochrome) in birds with LEMF. It was also based on the fact that this would disrupt other potential mediators capable of non-local interactions such as the π cloud of microtubules and membrane vibrational modes (potential mediators are discussed in the appendix). Under the *fifth postulate*, cells were permitted, when in culture plates, to grow to approximately 80% confluence. A confluence of 80% was chosen because it was felt it gave the maximum amount of opportunity to re-establish quantum correlations (again assuming close proximity is initially necessary) but did not allow growth to reach a point where substantial apoptosis began occurring spontaneously (near 100% confluence). The *sixth postulate* was that we needed to focus on testing the hypothesis first. "***is that one cell population can be altered in function by intervention on a second cell population non-locally without direct contact or biomolecule exchange." and it was consistent with quantum correlations.*** We knew that in testing the hypothesis, we would not be ruling out one alternate hypothesis that the non-local signaling was not due to intrinsic LFEF less than 100 MHz. Intrinsic LFEF signaling has never been described in humans and would require a considerable more complex experimental design to rule it out. This is consistent with a top down approach moving forward in a sequence of studies getting to the mechanism in steps. This will be addressed in more detail in the discussion.

IVG. Protocol:

The protocol for the HL-60 cell experiment is shown in figure 3. Individual vials of human promyelocytic leukemia cells (HL-60; ATCC, Manassas, VA) were thawed, suspended in Iscove's Modified Dulbecco's Medium with 20% fetal bovine serum, and plated into separate 25-cm$^2$ culture flasks to expand (humidified 5% $CO_2$, 37°C) to a cell density of approximately 1 X 10$^5$ cells/ml. Cells were intermittently agitated (under postulate 2). For each study, cells expanded from a single vial were transferred into 2 wells of a 6-well plate at a density of approximately 2 X 10$^5$ in fresh media. A concern in design was the potential the postulated correlations would be maintained between cell populations at the first split (third postulate) which would be undesirable at this split. So it was hypothesized that exposure to a strong LFEF would reduce the potential of these unwanted correlations for reasons described in the previous paragraph (fourth postulate). Cells were exposed to a magnetic field by rotating a ring magnet (H750L, Magnets, Irvine, Ca) 3" above the wells for 20 minutes at one rotation per second (max surface field 4640 G), where the field intensities are discussed in the results section. The magnet was brought to a complete stop after each 360$^0$ turn (1/sec) to create the LFEF.

Tissue culture plates were then returned to the incubator within plastic containers covered with magnetic shielding foil (0.004" thick, EMF Safety Superstore, Albany, NY) to at least provide minimal shielding from fields within the incubator that may interfere with re-establishment of correlations. But far more sophisticated shielding is needed to rule out the broad range of LFEF, which will be addressed in the discussion. The lids were loosely placed to permit ventilation. Cells were permitted to grow to approximately 80% confluence (fifth postulate). Cells were periodically agitated over this period under the fourth postulate that correlations would be reestablished by continued contact or close proximity.

In the next step, each well was resuspended in fresh media and split again evenly into 2 wells. Under the above postulates, based on the potential mechanisms, this split resulted in 2 pairs of wells (per run) with correlations anticipated within the pairs, but not between them. The four wells (per run) were $T^+$ (treated with an apoptotic agent to induce programed cell death) and $T^-$ (the quantum correlated partner treated with a vehicle alone) in the first pair; and, in the second pair, $C_1$ (control 1) and $C_2$ (the quantum correlated partner to control 1) both treated with vehicle alone. Apoptosis was induced in one well per run ($T^+$) by adding 40 uM Paclitaxel (Taxol; Sigma Co, St. Louis, MO) to the well while the remaining wells receiving vehicle alone (DMSO). Taxol is a microtubule-targeted tubulin-polymerizing apoptotic agent [118]. Following binding to B-tubulin, it inhibits microtubule dynamic instability and cell cycle $G_2$/M phase transition. This mitotic arrest of cancer cells triggers the molecular signaling for the mitochondrial pathway of apoptosis.

Cells were then incubated for 18 hrs followed by measurements of cell density using a Coulter counter. Caspase-3 and 7 activity (important mediators of end stage apoptosis) was also assessed using a Caspase Colorimetric Assay (Promega; Madison,WI) [138]. This is a luminogenic assay for caspase 3 and 7. The caspase activity is measured via a luminometer (Wallac Victor3 1420, Perkin Elmer, Waltham, MA). Results are expressed as total maximum luminosity / cell count. Experiments were performed by a blinded investigator to the theoretical basis of the experimental design. The wells were randomized with respect to relative position and repeated for n=5.

Data Analysis:

Both cell numbers and caspase activity are represented as means ± standard error. One tailed paired t-tests were performed with the assumption that the results are sampled from a Gaussian distribution. Significance is defined as p<0.05. T¯ test were performed using InStat (GraphPad Software Inc). The power analysis was performed with JAVA applets (www.cs.uiowa.edu).

Results

The strength of the field without rotations was 230 to 355 Gauss (DC periphery to center) at 3 inches above the wells. With the rotations the AC the oscillation of the LFEF was 2 mG to approximately 1000 G during the rotation (Trifield Meter Model 100 XE and Gaussmeter model 1, Alphalabs, Salt Lake City, Utah). The field was applied after splitting into the T and C groups to disrupt potential correlations between them. As the separated cells were allowed to grow in culture, it is postulated correlations were then re-established within the T and C groups (as per postulates 2).

The results of the HL-60 experiments are shown in figure 3. The average cell counts in the control group (average between $C^+$ and $C^-$) were $2.1 \times 10^5 \pm 1.7 \times 10^4$ at the end of the experiment (no significant difference between control groups, below). Average cell counts in the $T^+$ group were $1.3 \times 10^5 \pm 2.8 \times 10^4$ while in the $T^-$ group they were $1.2 \times 10^5 \pm 1.9 \times 10^4$. There was no significant difference between the $T^+$ and $T^-$ groups, consistent with the hypothesis of non-local signaling, which we postulate was quantum correlations. The p value with respect to $T^+$ versus $T^-$ cell counts was 0.35 (NS). The difference between the $T^+$ and controls groups was highly significant (p < 0.001) while more importantly the difference between the $T^-$ and controls was p < 0.01. <u>The difference between the C and $T^-$ groups confirms non-local signaling in combination with no difference between $T^+$ and $T^-$ groups.</u> There is no previously described mechanism for this and it is at least consistent with quantum correlations. In other words, mortality in the $T^-$ groups should have been the same as controls, but instead was equivalent to the $T^+$ group. There was no significant higher difference between the control groups with $C^+$ being $1.9 \times 10^5 \pm 2.6 \pm 10^4$ and C- was $2.2 \times 10^5 \pm 1.4 \times 10^4$ (p = 0.16, NSD) which was expected. The power analysis is 1.0 under the assumption of p=0.05 for a one tailed paired t-test with an observed effect of 10%.

The average caspase activity was $0.32 \pm 0.06$ for the $T^+$, $0.17 \pm 0.02$ for the $T^-$, and control $0.098 \pm 0.009$. The p value between the $T^+$ group and controls was < 0.01, between the $T^-$ and controls was < 0.01, and between $T^+$ and $T^-$ was < 0.05. Therefore, there was no significant difference in cell mortality between $T^+$ and $T^-$, but there was between their caspase 3/7 activity, suggesting nonlocal cell death is not exclusively through a caspase mechanism. However, the $T^-$ still had statistically significant higher caspase activity compared to controls. The $C_1$ and $C_2$ groups were $0.1 \pm 0.01$ and $0.09 \pm 0.005$ (P= 0.2673, NSD). The power analysis is 1.0 under the assumption of p=0.05 for a one tailed paired t-test with an observed effect of 10%.

V. Discussion

The driving of large-scale systems by quantum rather than classical mechanics has been pursed for over a half century because of its potential for paradigm shifts across a wide range of fields. If the extreme efficiencies achieved with mesoscopic/ macroscopic quantum subsystems in photosynthesis and magnetoreceptor can be accomplished with other subsystems/systems, it would suggest the impact of utilizing MQS (intrinsic or created) seems virtually unlimited. Ignoring for now their potential for advancing man-made devices, it is unclear how many other biological quantum subsystems already exist undetected within the otherwise complex classical environment. As we have demonstrated and will review below, the major challenge for both nature and engineering isn't creating MQS. The difficulty exists in maintaining them in the setting of the heavy environmental entanglements (decoherence) of complex systems. We have very little insight about how this is achieved. Reductionist theories give us insights into simple systems far from ambient conditions. From a reductionist standpoint, the assumption would be that results on a small scale could be extended to more complex systems. So as quantum systems get larger, the number of interactions increases and therefore larger systems are doomed to failure from decoherence. But the focus of this paper is that this is an invalid assumption, where extrapolating reduction models from smaller to larger systems is not viable for the reasons discussed and reviewed below. We actually have little insight into how decoherence is controlled in mesoscopic/ macroscopic quantum subsystems.

Creating a MQS is theoretically not a challenge, but maintaining it is. A critical point of the paper is a coherent system remains coherent unless it interacts with an outside system (i.e. closed coherent systems remain coherent), so without interaction it is sustained. This was shown with the progression of the density operator under unitary transform without environmental interaction, a closed system (it was also illustrated with a simple Young's interferometer). Also using the simple example (a reductionist experiment) of a Young's interferometer and then with density operator formalization, we demonstrated that environmental entanglements generally result in coherence loss. But under the right conditions (not generally dealt with in reductionist approaches) some environmental entanglements in even simple systems lead to the state being unaffected, a new small coherent state, or expansion to a larger coherent system (closed system to larger closed system). Even these simple coherence preserving conditions could play a role in decoherence compensation in complex systems, but the situation is far more complicated.

Beyond this simple analysis of decoherence, where there are even instances in which environmental interactions do not result in decoherence, we demonstrated reductionist analysis is inadequate for modeling mesoscopic/ macroscopic quantum subsystems. Most importantly the extension of reductionist results to complex systems ignores higher order emergent behaviors such as compartmentalization, robustness, and broad adaptability that are lacking in simpler systems [47-59]. Not just the system but the environment must be treated as having a potential role in sustaining coherence. In addition, as discussed, the reductionist approach does not take into account additional factors (beyond emergent behavior) that need consideration when examining complex systems. These additional factors do not allow calculations from master equations to be extrapolated to complex systems. The first of these is the essentially universal use of approximations (Born and Markov most commonly) in decoherence master equations that would not be valid for CAS. These approximations essentially treat the environment as an infinite homogeneous pool and not that it can, for example, become part of the coherent system as in the discussed diamond study. The second is not taking path integrals into account in master equations, which may preserve coherence in a complex system. The third is not addressing decoherence avoidance or compensation mechanisms, even those already demonstrated with man-made systems (such as in the quantum information sciences). The fourth is not addressing the subentity problem, particularly in the use of the trace of density and time evolution operators in master equations. Therefore, reductionist master equations can't represent complex systems, so the difficulty in demonstrating MQS using experimental designs based on this reductionist modeling is not surprising.

As mechanisms preventing decoherence can't be identified by extending results of reductionist experiments to larger systems, experiments need to be performed top down. It requires working with as intact of a system as possible and basing experiments on educated postulates from prior results or observations. Models are gradually constructed, which can then be ultimately refined to understand more detailed aspects of the system. This maintains large-scale behavior, such as the emergent behavior of CAS. But by analyzing from top down, the other factors listed that are not addressed with decoherence master equations are incorporated in the top down modeling. This is achieved by finding equivalent classes and transition functions that result in the development of homomorphisms. These equivalent classes and transition functions are obtained examining the system, as intact as possible (with just sufficient reductionism to identify responses), and assessing responses to varying stimuli. There are many successful examples of this top down approach in science from the progression of Mendelian genetics to photosynthesis and magnetoreception, though all these fields progressed about 100 years before identifying the primary mediators. This is the core point of the paper.

For this paper, a top down experiment was performed to illustrate the approach and examine it for a potentially new mesoscopic/macroscopic quantum subsystem. A biological system was chosen for several reasons already described, but predominately because it is a CAS, which has been developed/refined by evolutionary forces. Biological systems were also chosen because they are again known to sustain mesoscopic quantum systems operating under ambient conditions.

We performed a study evaluating the potential of non-local (remote) quantum correlations between leukemic cell populations (HL-60 cells). The experiment was based first on anecdotal correlations between HL-60 cells where direct contact or mediator exchange did not appear possible. It was also supported by work from ours and other groups demonstrating macroscopic quantum correlations with optical systems. Finally, it was supported by our observations on the limitations of reductionist analysis in experimental design (the focus of this paper). We believed that these reasons, along with those already described, would lead to a higher likelihood of success, which was the case.

The ultimate design then was based on the hypothesis that remote signaling occurred which was not mediated by contact or biomolecule exchange (and <u>consistent</u> with quantum correlations). The top down experiment does achieve this as will be described, but as is common with top down experiments, additional work needs to be done to further support the mechanism for reasons discussed below (top down studies are a progression). Briefly, in the experiment, apoptosis was induced in one leukemic cell population and death occurred in a spatially separated paired set of cells receiving only vehicle. The postulated non-local communications were believed secondary to nonlocal quantum correlations in apoptotic machinery between cells (consistent with anecdotal observations and success in non-locality optical experiments). Furthermore, it was postulated that these quantum-correlated subsystems could be sustained and manipulated. So this assumes no correlations between the T and C groups existed after splitting due to exposure to a high intensity LFEF. In addition, in accordance with a second postulate, if cells were allowed to grow, correlations were re-established <u>within</u> groups (T or C). Cell death rates were not statistically different between the T+ (treated with a apoptotic agent Taxol) and T- groups (without Taxol) but were statistically different from the controls. Both the lack of difference between the T groups and the statistical difference between the T- and control groups would be consistent with non-local quantum correlations. As stated previously, the results of these experiments do not likely imply that if the quantum correlations do exist, they are between entire cells. We feel it is more likely that mesoscopic systems within the cells are correlating non-locally (even if the underlying mechanism is LFEF), but still affecting overall cell function (we use the radical spin of cryptochrome as an example mediator). This would be analogous to the magnetoreceptors, a subsystem, controlling whole bird flight patterns. It is also again consistent with the concept that decoherence and coherence need to (and do) exist in close proximity.

The study also showed that the T groups had statistically higher caspase activity than controls, again consistent with non-local signaling. However, the caspase activity was statistically lower in the T- cells compared to the T+ cells, suggesting that the mechanism of cell loss between the two T populations may not be identical. It is likely the T- cells at least in part are operating through a caspase 3/7 independent pathway based on the results of figure 4. This could possibly be apoptosis inducing factor (AIF) that in HL-60 is regulated by cryptochrome [141].

The experiment established that remote signaling occurs between HL-60 cell lines not mediated by direct contact or biomolecule exchange. But while the top down experiment answered this, it did not exclude at least one other mechanism (besides quantum correlations) that will be the source of future investigation for reasons discussed. An alternate (though less likely) possibility to quantum correlations is classical intrinsic LFEF, which have not been described in humans and would be extremely difficult to detect or filter. Though they are clearly generated by structures like the brain and microtubules, a role in signaling has never been demonstrated.

The idea that animals can even detect Earth's extremely weak magnetic field (LFEF) has gone from being ridiculed to a well-established fact in a little more than one generation. A plethora of experimental data, almost exclusively top down, has now shown that diverse animal species, ranging from bees to salamanders to sea turtles to birds, have these internal compasses. The 50 µT LFEF of the Earth is six orders of magnitude below thermal noise, contributing to the initial skepticism it was used in migration. Studies have shown the magnetoreceptor, now strongly believed to be cryptochrome, can be influenced by LFEF even in the nT range. Fields this weak could be used in non-local intrinsic signaling (including correlating in cell mortality), which would be very difficult if not impossible to measure directly in complex biological environments. So very weak fields can effect biological systems and go completely unobserved. These classical fields are produced by electron oscillations, for example, by membranes (outer or internal) or microtubules [139-149]. An example molecule that can detect fields in this range is cryptochrome that works through a radical pair mechanism (appendix). Cryptochrome is present in neutrophils/HL-60 cells (though the function of it's field sensitivity is unknown) where it is a mediator in apoptosis, the cell cycle, and the peripheral circadian clock.

The current study did achieve its objective in demonstrating the top down approach and identifying non-local signaling that was independent of contact and biomolecule exchange, consistent with quantum correlations. But it did not exclude the possibility of previously undescribed LFEF because adding this additional arm to the experiment requires removing or altering LFEF. LFEL are difficult to block, whether it is being done to remove intrinsic signaling between cells or to remove ambient fields. This is why we choose to focus on demonstrating that the non-local signaling exists and once demonstrated, we will now exclude weak LFEF with the next series of top down experiments. Frequencies under 100 Hz can penetrate 10 feet of concrete with negligible attenuation. In future studies, for these frequencies we will use physical distance (the field decreases in intensity at $1/r^2$) or 80% nickel diffraction gratings (which randomizes the signal but don't attenuate it) [142]. For frequencies above 100 Hz and below 100 MHz, dense aluminum or copper can be used to block the signal [142]. Being able to block fields between cultures is critical to studying mechanisms when you can't directly measure them. Simple magnetic foil or lead/concrete blocks are insufficient.

The experimental portion of the paper explores the possibility of non-local correlations between subsystems in cells. The experiment was included primarily to demonstrate the principles of the top down experiment, confirm that non-local signaling exists (as seen in anecdotal studies), and that it is consistent with quantum non-local correlations. This was achieved. As local realism has fallen out of the mainstream at this point, the discussion should be focused on can decoherence compensation allow quantum correlations to be sustained for this period of time. Discussing whether non-local correlations exist should be simply accepted at this point. However, because non-local quantum correlations are still counterintuitive to even physicists, resistance to the possibility of non-local correlations between remote cell subsystems (even in the absence of decoherence) remains high. After the extensive discussion above, the challenge is preventing coherence breakdown as there is no law in quantum mechanics preventing non-local quantum correlations between subsystems. It is not the intention of this paper to provide a justification for a nonlocal reality, that already has been established [143-144]. But much of the 'paradoxes' of quantum mechanics, a function of a classical view, disappear when reality is accepted as being primarily nonlocal and deterministic and then becoming nondeterministic and local with measurement. To paraphrase recent statements by Vedral and Hawkings, among others, experimental data is pointing toward classical (including relativistic) space-time as secondary to a non-local reality [143-144]. The most challenging aspect of the current experimental results is the duration of the non-local correlations, which appears to have extended over several days, and not the non-locality, per se.

Potential mediators of quantum correlations are discussed in the appendix. This is a top down experiment, so as with the analogies of Mendelian Inheritance and photosynthesis, it is expected that establishing events occurring on the microscopic level will likely take a considerable number of studies. Therefore, discussions on mediators in the appendix are purely speculative and only intended to illustrate that there are subsystems that have the appropriate characteristics.

**VI.** Conclusion

The paper focused on the limitations of current reductionist approaches to decoherence for complex systems. Results found in small systems under extreme conditions can't be simply extrapolated to more advanced systems under ambient conditions. Specifically, complex adaptive systems (CAS) are not amenable to reductionist models (and their master equations) because of emergent behavior, approximation failures, not accounting for quantum compensator mechanisms, ignoring path integrals, and the subentity problem. <u>This paper advances the mesoscopic/macroscopic quantum field by a complex systems approach to decoherence</u>. We argue we know very little about compensatory mechanisms in CAS and their study requires a top down approach. Furthermore, biological systems likely offer the greatest opportunity for studying complex decoherence as they have benefited from evolutionary development to use approaches currently not conceived.

This is exemplified by the extremely high efficiencies seen in the quantum subsystems of photosynthesis and magnetoreception.  In addition, a study was included which illustrated the top down approach with an experiment on non-local signaling between cells which neither involves cell contact nor biomolecule exchange.  The results did demonstrate the effectiveness of the top-down approach and that the non-local signaling exists.  Furthermore, the results were consistent with non-local quantum correlations.   But further study is needed to rule out classical LFEF, a mechanism not previously described in humans.  In response to the question we posed in the title "Can we advance macroscopic quantum systems outside the framework of complex decoherence theory?", we believe we have answered the question no.

**Acknowledgments**:
This work was supported by Dr. Brezinski's funding from the National Institute of Health Grants R01 AR44812, R01 HL55686, R01 EB02638/HL63953, R01 AR46996, R01 EB000419, and R21EB015851-01. The authors would like to thank the efforts of Bin Liu PhD and Chris Vercollone in preparing the paper.  We would also like to thank Joseph Loscalzo MD, PhD for his insights.

**Conflict of Interest:**
The authors report no competing interests, financial or otherwise, with the publication of this material.

**Appendix 1: Mesoscopic Quantum Biological Systems.**

This appendix discusses two mesoscopic quantum systems, magnetoreception and photosynthesis raised in the main text. Also mentioned are potential mediators. This is a top down experiment, so as with the analogies of Mendelian Inheritance and photosynthesis, it is expected that establishing events occurring on the microscopic level will likely take a considerable number of studies. Therefore, discussions on mediators in this appendix are purely speculative and only intended to illustrate that there are subsystems that have the appropriate characteristics.

1A. Magnetoreceptors

This approach to non-local correlation in this paper has similarities to magnetoreceptor work as it is top down, likely will face considerable skepticism (as it is paradigm shifting), and the ultimate mechanism can be disrupted by LFEF. So magnetoreception is discussed here. Magnetoreception is a sense that allows an animal to detect a magnetic field to perceive direction, altitude or location. Magnetoreceptors are particularly relevant here because many, if not most, operate through a mesoscopic quantum system. Specifically, they work through the radical pair mechanism in the molecule cryptochrome (found in the bird's eye). These systems have the ability to detect the gravitational fields, six orders of magnitude weaker than thermal noise, making them better than any classical man made device.

The idea that animals can detect Earth's magnetic field, below 50 µT, has gone from being ridiculed to a well-established fact in a little more than one generation. A plethora of experimental data has now shown that diverse animal species, ranging from bees to salamanders to sea turtles to birds, have internal compasses. A series of top-down experiments, needed because the CAS would be destroyed by a reductionist approach, gave insight into the mediators. These included the fact magnetoreception required blue light (ineffective with red), could be disrupted to by a 1.4 MHz field in the nT range, the inclination of the earth's magnetic field altered migration, and blindfolded robins did not respond to a magnetic field at all. These types of top down experiments led to the identification of cryptochrome as the magnetoreceptor, with two unpaired electrons that form an entangled pair with zero total spin (cryptochrome is discussed under mediators). When this molecule absorbs blue light, the electrons get enough energy to separate, either into a singlet or triplet state, and become susceptible to external influences, including the earth's magnetic field. Chemical pathways in the eye translate this difference into neurological impulses (an example of the interface between quantum and classical systems). If one assumes that the radical pair in the triplet state forms a chemical product that differs from that of singlet pairs, the ratio of which is determined by the external field, one has a potentially viable detector for weak magnetic fields.
1B. Photosynthesis

Photosynthesis is of interest in this paper both because, like magnetoreceptors, the underlying mechanism is a mesoscopic quantum system and the field progressed forward top-down. Photosynthesis has been studied for several hundred years and investigations have their origins in a top-down approach, starting with water and carbon dioxide (in soil and air) utilization in plants in the 18th and 19th century. Whole plants were studied initially in response to various environmental changes. Eventually, the mediator was determined to be the green absorbing pigment chlorophyll, though mechanistic specifics required further study. It became clear classical mechanics did not account for the near perfect efficiency of photosynthesis. In the older classical model of photosynthesis, in what is known as the Foster model, the electrons would 'bounce' around the light harvesting-storing complex until they found a center that was ready for absorption. This process would be highly inefficient if nature did not provide a quantum mechanical alternative. The actual mechanism is a non-classical non-local superposition of electrons and phonons (excitons), which achieves extremely high efficiencies. For illustrative purposes, an electron 'blanket' exists over reaction centers, some available or some not. Where available reaction centers exist, they pull on the closest electrons in the blanket, bringing other electrons in the 'blanket' with them (due to the quantum non-locality). So after well over a century of top down experiments refining understanding, reductionist studies were then used. For example, ultra-short pulsed lasers in combination with heavily isolated photosynthetic units under extreme conditions (reductionist) were used. These demonstrated the quantum nature (delocalized electrons) of the photosynthetic unit. So, these reductionist approaches become more important once extensive top down data has been produced. With photosynthesis, these reductionist studies have been generally done with isolated light-absorbing units under extremely low temperatures. But, even in the setting of these experiments which use a reduced number of elements with far from ambient conditions, unexplained low decoherence rates have been found than are predicted from master equations. As they are much less than expected, the suggestion has been made that 'shielding' is occurring from local proteins within the unit, which is a classical attempt at an explanation.[19] A more likely possibility is that some residual emergent behavior remains, as the complex quantum system is partially intact, which reduces decoherence rates. This concept of shielding has, in the author's opinion, haunted the field of quantum biology and appears in the photosynthesis, magnetoreceptor, and microtubule literature to explain decoherence rates lower than anticipated. This further emphasizes the need to treat decoherence from a complex systems approach. The main text discusses how decoherence compensation mechanisms for CAS are undoubtedly more advanced then those envisioned for reductionist systems and would not require the classical 'shielding' concept. Looking again at a CAS analogy, skin being continuously shed by living creatures, there is no need to invoke a protective shield to prevent skin loss or that the organism will die because of it. The CAS regenerates through higher order behaviors (both the skin and the organism) that is far from simple but preserves the system (the organism).

**Appendix 2: Mediators**

The focus of the text is the need for a complex systems approach to decoherence, primarily top down, and the limitations of reductionist decoherence approaches. In addition, a study is included which illustrated the top down approach with an experiment on non-local signaling between cells which neither involves cell contact nor biomolecule exchange. The results did demonstrate the effectiveness of the top-down approach and that the non-local signaling exists. Furthermore, the results were consistent with non-local quantum correlations, though did not exclude LFEF. Commenting on the mediators of this study at this stage, as with most early top down studies, is highly speculative. It is done simply for the purpose of demonstrating that potential mediators exist. We will discuss cryptochromes, in particular, as potential candidates as well as microtubules.

2A. Cryptochromes

We believe cryptochromes are the most likely mediators of our non-local observations, whether the correlations are quantum mechanical or even if through LFEF. Cryptochromes are a ubiquitous class of flavoproteins found in plants and animals. They are now universally held to be the magnetoreceptor for migratory animals. In plants they also guide growth. In addition, they are critical to the master circadian clock in humans, which is located in the suprachiasmatic nuclei (SCN) in the hypothalamus. Details on this can be found elsewhere so it will not be discussed as it is not directly relevant to the paper. But of substantial importance to this paper, it is a major mediator of the cell cycle, cellular/tissue circadian rhythm, and apoptosis.

With respect to apoptosis, cryptochrome appears to work through a caspase independent mechanism, possibly apoptosis inducing factor (AIF). This may explain in part the apoptotic markers results seen in this study. However, an intriguing question is why the radical pair, which can sense magnetic fields nine orders of magnitude below thermal noise, is needed in inflammatory cells. These roles of cryptochrome are LFEF sensitive.

Looking at why crytochromes are an example of potential mediators for the results of the current study for several reasons:

1. They have been shown to drive mesoscopic quantum phenomena in other systems, particularly magnetoreception. In many organisms, they also are an important component of the photosynthesis quantum machinery.
2. They are present in HL-60 cells and neutrophils in substantial concentrations.
3. They regulate apoptosis in many cells including white cells.
4. They are cell cycle mediators.
5. They are circadian clock mediators, both the master and peripheral clocks.
6. They are the most sensitive biomolecules to external fields.
7. They are linked to the neoplastic transformation in leukocytes.
8. Correlation of the radical spins offers a mechanism for nonlocal correlations.
9. Their function is altered by LFEF.

Cryptochrome's extreme sensitivity to magnetic fields is due to the presence of this radical pair. Radicals are molecules that have an odd number of electrons and consequently an unpaired electron spin that may be found in one of 2 spin states: ↑ or ↓. A radical pair is an intermediate consisting of 2 radicals formed in tandem whose unpaired electron spins may be either antiparallel (↑ ↓, a singlet state, S) or parallel (↑ ↑, a triplet state, T). As each electron spin has an associated magnetic moment, the interconversion and chemical fates of the S and T states can be influenced by internal and external magnetic fields. In chemical terms, the minimum requirement for a radical pair reaction to be sensitive to an external magnetic field is that at least one of the S and T states undergoes a reaction that is not open to the other. The radical pair is formed by the absorption of blue light in magnetoreception, but this state can also be achieved by either a local field or an oxidative reaction. Hyperfine interactions are crucial because they drive the interconversion of the S and T states of the radical pair and allow it to be modified by an external electromagnetic field. In particular, this external magnetic field has a strength near 30-80 $\mu$T and a oscillatory frequency of 1-10 Hz. S $\iff$ T interconversion is a coherent quantum mechanical process: radical pairs oscillate between their S and T states at a variety of frequencies determined by the strengths of the hyperfine interactions. This is the resonance frequency of the spin precession. The spin precessions, in organic radicals, are in the range 10–1,000 $\mu$T, corresponding to frequencies of 300 kHz to 30 MHz. So three fields are involved in cryptochrome function, the blue light (approximately 550 nm) or equivalent energy to a blue light photon, the magnetic field (under 10 Hz), and field altering the precession frequency (100 kHz to 100 MHz).

2B. Microtubules:

In addition to cytochromes being a potential mediator for non-local signaling seen with the HL-60 cell study in this paper, several other intermediates also seem to be feasible. In particular, microtubules will also be considered as potential mediators of the observed signaling.

Microtubules constitute a major part of the structural network, or cytoskeleton, of cells. They are fibrous, hollow rods that function to support and shape the cell as part of the cytoskeleton, as well as play a critical role in movement. They also function as routes along which organelles can move. For example, they form the spindle fibers that manipulate and separate chromosomes during mitosis, as well as play a role in the cell cycle and apoptosis.

Microtubules are made up of tubulin protein subunits, which is what makes them attractive for establishing correlations through quantum correlations or even LFEF. The tubulin protein dimers of the microtubules have hydrophobic pockets that contain delocalized π electrons, important to either large scale quantum systems or electron oscillations. Tubulin has other smaller non-polar regions, for example 8 tryptophans per tubulin, which contain π electron-rich indole rings distributed throughout tubulin with separations of roughly 2 nm. Hameroff claims that this is close enough for the π electrons to become quantum entangled.

Several other relevant points about microtubules that have been made by Hameroff and Penrose:
1. Microtubule individual subunit (tubulin) conformations may be coupled to quantum-level events (electron

movement, dipole, phonon) in hydrophobic protein regions.
2. Microtubule paracrystalline lattice structure, symmetry, cylindrical configuration, and parallel alignment promote long-range cooperativity and order.
3. Hollow microtubule interiors appear capable of water ordering, waveguide super-radiance, and self-induced transparency.
Future work is needed in this area but microtubules are capable of sustaining quantum correlations as well as generating and receiving LFEF.

## 2C. Other Potential Mediators:

Cryptochrome and microtubules are potential mediators for the non-local signaling observed, but others have also been proposed, such as the vibrational modes of membranes. As this section is purely speculative to illustrate the potential, we will limit the discussion to what has already been presented.

**Figure Legends**:

Figure 1. Simple Coherence and Decoherence Model. In this figure, decoherence is demonstrated in terms of loss of path indistinguishability from local environmental entanglement with a Young's interferometer.

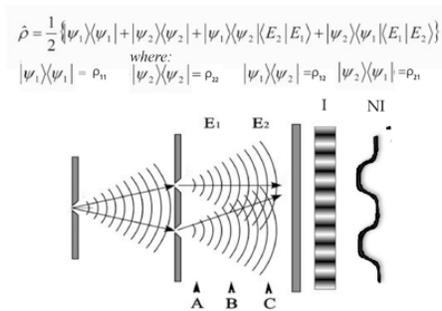

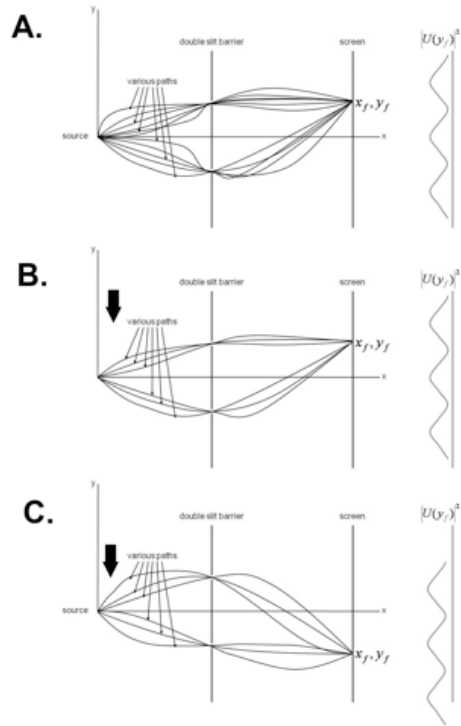

Figure 2. <u>Decoherence Model Accounting for Path Integrals</u>. Unlike the previous figure, which only looks at the action, here a limited description of decoherence with respect to path integrals is provided. In figure 2A, Young's experiment is depicted, except with a partial path integral (12 paths rather than an infinite number for illustration) rather than a single path. Note that the path integral for the Young's experiment is described from the source to the screen involving both slits (i.e. the action involves both slits). There are no decoherence/environmental entanglements in figure 2A, so an interference pattern exists. Now in figure 2B, the action and interference pattern are maintained in spite of a reduction of some paths from decoherence. In figure 2C, coherence and interference are present, but altered, as the action has been phase-shifted. Therefore, destruction of some paths does not necessarily correspond to loss of coherence or the action.

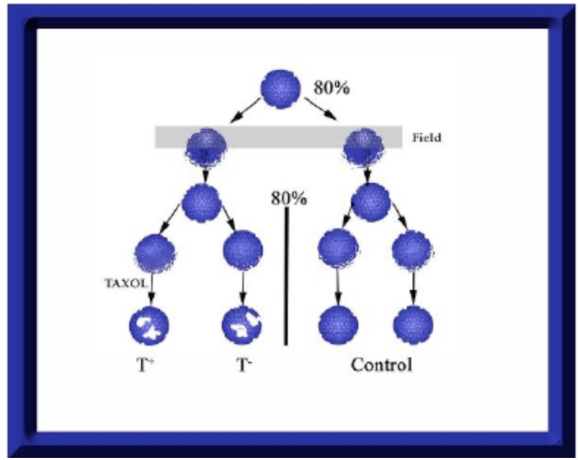

Figure 3. Protocol for HL-60 experiment. The experiment intends to confirm the non-local signaling identified anecdotally and also illustrates the top-down approach. T⁺ represent the Taxol (apoptosis) treated group. T⁻ (treated with vehicle ) represents the cells theoretically entangled to the T⁺ group. Control represents vehicle-treated cells not correlated with either T⁺ or T⁻ cells. Arrows represent cell splits (angular) or growth (vertical) while the gray area is low frequency high intensity EMF exposure.

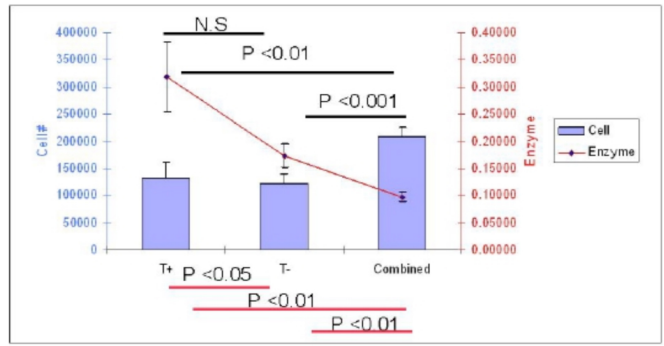

Figure 4. <u>Experimental Results in Terms of Cell Counts and Caspase Activity</u>. $T^+$ and $T^-$ cells shows no significant difference in cell mortality, consistent with a quantum mechanical effect. Similarly, the T- group had a statistically reduced cell number relative to the controls, consistent with a quantum mechanical effect (but not proof). With respects to the caspase activity, it was significantly higher in T groups than control. The lower activity in $T^-$ relative to the $T^+$ groups suggests cell death in the $T^-$ group was not exclusively through the caspase mechanism.